\PassOptionsToPackage{left=3cm, right=3cm, top=3cm, bottom=3cm}{geometry}

\documentclass[pdflatex,sn-mathphys-num]{sn-jnl}

\usepackage{graphicx}%
\usepackage{multirow}%
\usepackage{amsmath,amssymb,amsfonts}%
\usepackage{amsthm}%
\usepackage{mathrsfs}%
\usepackage[title]{appendix}%
\usepackage{xcolor}%
\usepackage{textcomp}%
\usepackage{manyfoot}%
\usepackage{booktabs}%
\usepackage{algorithm}%
\usepackage{algorithmicx}%
\usepackage{algpseudocode}%
\usepackage{listings}%

\theoremstyle{thmstyleone}%
\theoremstyle{thmstyletwo}%

\theoremstyle{thmstylethree}%

\begin{document}

\title[Article Title]{Adapting ILC detector concepts to other facilities}


\author{\fnm{Daniel} \sur{Jeans}} \email{daniel.jeans@kek.jp}




\affil{\orgdiv{IPNS}, \orgname{KEK}, \orgaddress{\city{Tsukuba}, \country{Japan}}}




\abstract{
  Detectors designed for the International Linear Collider have been studied over many years,
  and a good understanding of the relevant requirements and constraints has been developed.
  With the prospect of future Higgs Factory projects with different properties (for example, the FCC-ee),
  the adaptations required for use at such facilities are being investigated.
  We outline our current understanding in this note.
}


\maketitle

\section{Introduction}\label{sec1}

A number of detector concepts have been developed for use at the International Linear Collider (ILC) \cite{Behnke:2013xla, Abe:2025yur}, most
notably the Silicon Detector (SiD) \cite{Breidenbach:2021sdo} and the International Large Detector (ILD) \cite{ILDConceptGroup:2020sfq}.

Over the last years, proposals for an electron-positron Higgs Factory collider are advancing in Europe,
with CERN as the assumed host lab. In this note we will emphasise mostly the electron-positron Future Circular Collider
(FCC-ee) \cite{FCC:2025uan, FCC:2025lpp} and the Linear Collider Facility (LCF) \cite{LinearCollider:2025lya}.
The Circular Electron Positron Collider (CEPC) \cite{CEPCStudyGroup:2023quu, CEPCStudyGroup:2025kmw} is in most respects very similar to the FCC-ee. 
Of the other major proposals, LEP3 \cite{Anastopoulos:2025jyh} can from the detector point of
view be considered as a scaled-down version of the FCC-ee, while the electron-proton collider LHeC \cite{Ahmadova:2025vzd} will likely
require a significantly different design.

A number of important differences between these facilities need to be taken into account, in particular the
design of the Machine-Detector interface, including details of the final focus system and associated magnetic fields,
different ranges of collision energy, and the varying collision rates at different energy points.

A similar detector design progression has been followed in the transition from the ILC detectors to the CLIC detector design \cite{Adli:2025swq, CLICdp:2018vnx},
to the CLD concept \cite{Bacchetta:2019fmz} for FCCee.

\section{Physics requirements}

The requirements on detector performance at ILC and the other proposed electron-positron Higgs Factories have many similarities,
built around a run around the ZH threshold $240\sim250$~GeV which plays a prominent role in most projects.
It is the measurement of the recoiling Z in the Higgs-strahlung process, particularly in its muonic decay, which places the most
stringent requirement on the charged particle momentum measurement. The width of the recoil mass distribution, ideally
just the Higgs decay width, is broadened by the collider's beam energy spread. The momentum resolution of the muons should not 
add a dominant contribution to this width. It is this consideration which leads to momentum resolution requirement of $dp/p^2 \sim 2\times10^{-5}$.

A natural goal for the jet energy resolution is to enable the effective separation of hadronically decaying W, Z, and Higgs bosons by means of their di-jet mass,
which requires a di-jet mass resolution similar to or smaller than than the natural relative widths of the W and Z bosons.
Higgs-strahlung with a hadronically decaying Z boson -- the statistically dominant decay channel -- also argues for as good a di-jet four-momentum resolution as possible.

The ability to identify heavy flavour jets initiated by b or c quarks drives the need for precise reconstruction of tracks' impact parameter.
Being able to reconstruct the flavour of single particle (i.e. the different species of lepton and hadron) is of importance for many analyses, including
the reconstruction of specific decay channels of various states, and the correct accounting of particle masses.

Circular and linear colliders emphasise different energy ranges beyond this central Higgs-strahlung program.
Linear colliders, with their extended energy reach, require detectors designed to operate well at higher centre-of-mass
energies, for example measuring higher energy single particles (both neutral and charged) and hadronic jets with sufficient precision.
To maintain the effective separation between highly boosted hadronically decaying W and Z is a key challenge.

The lower energies emphasised by circular colliders, with their high statistics runs at the Z pole and WW threshold, place
different emphasis on detector design. Jets are generally less boosted, so confusion in particle flow reconstruction is less critical.
The high statistics of the Tera-Z run promise the search for rare decays in flavour physics, which will profit significantly from
enhanced particle identification abilities, for example between different hadron species.
The extremely small statistical uncertainties promised by a Tera-Z program should not be dominated by systematic uncertainties
due to the detector properties. While the same data will provide many opportunities for data-based calibration to control systematic effects,
as much robustness as possible should be designed into the detector.

These various considerations lead to a multi-dimensional optimisation problem in detector design.
Optimal regions in this space differ for detectors at the different future colliders.


\section{Machine Detector Interface}\label{sec2}

The design of the Machine Detector Interface (MDI) is in many respects defined by L*, the distance from the interaction point (IP)
to the last focusing quadrupole magnet of the beam delivery system, and the crossing angle of the beams at the IP.
These two quantities are related, since the space around the beampipes created by the crossing angle must be large
enough to house the magnets needed to focus and steer the in- and out-going beams.

\begin{figure}
  \begin{center}

    \includegraphics[width=\textwidth]{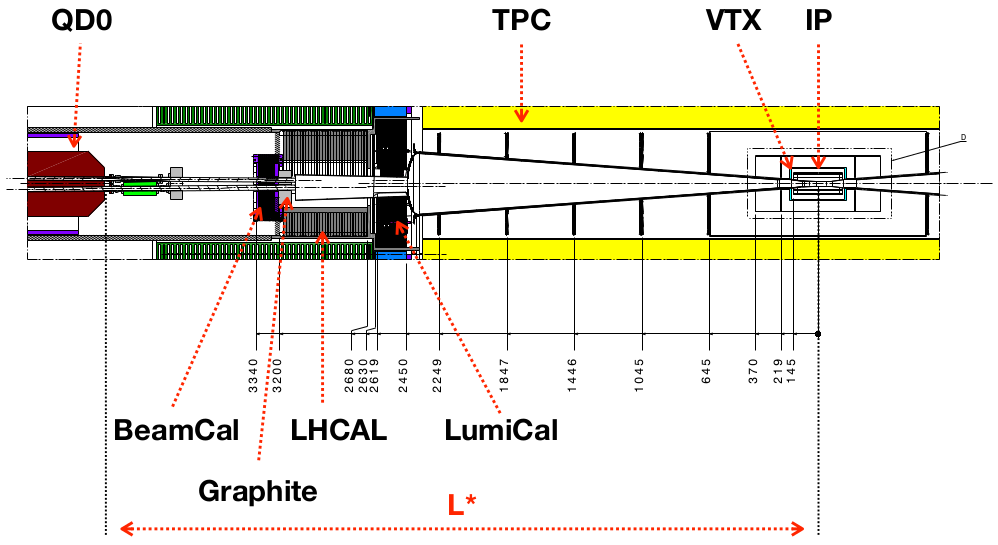} \\

    \includegraphics[width=\textwidth]{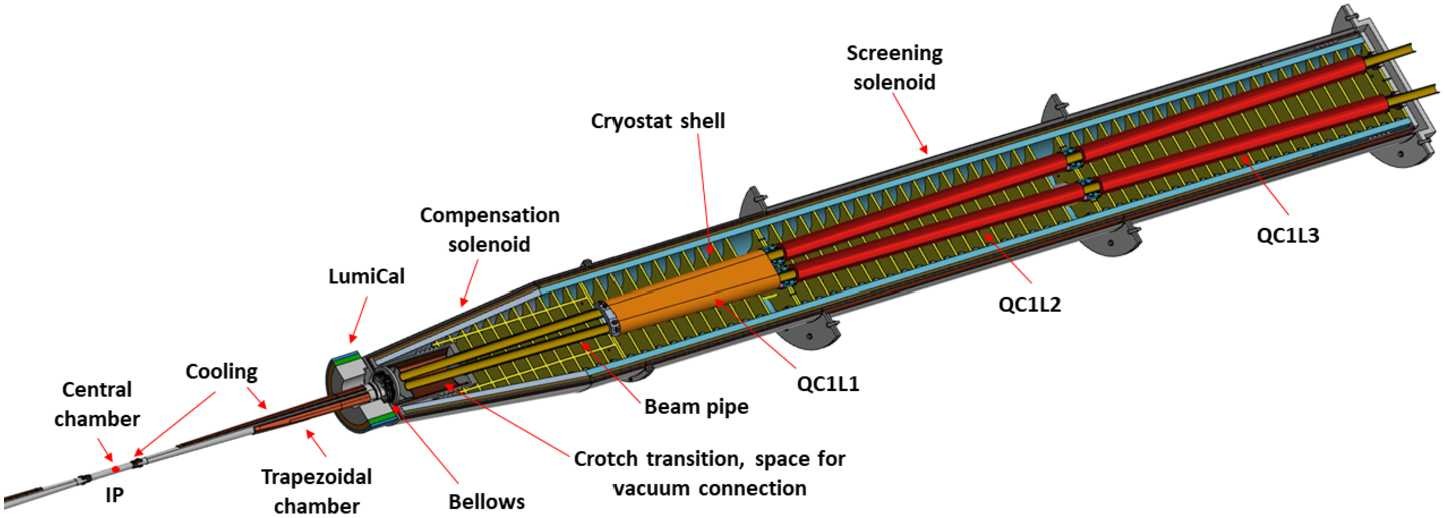}

  \caption{Mechanical drawings of the MDI systems at ILC (top, from \cite{ILDConceptGroup:2020sfq}) and FCCee (bottom, image from \cite{Boscolo:2023hwo}).}
  \label{fig:fccmdi}
  \end{center}
\end{figure}

At the ILC, a common L* of 4.1~m was agreed between the machine and detector concepts, at a crossing
angle of 14~mrad. This allows all accelerator-related components to be situated outside the
detectors' tracking region.
The different requirements of the FCCee lead to a MDI with a notable shorter L* of 2.2~m
together with a larger crossing angle of 30~mrad \cite{Boscolo:2025ntr}.
The current design is shown in Fig.~\ref{fig:fccmdi},

The design of the LCF MDI is not yet fixed, beyond an assumption that two interaction regions will be present.
The choice of crossing angles at the two IPs, which is baked into the facility design through large-scale civil construction,
will likely depend on constraints from the machine side, including those from
potential future upgrades of the facility (for example to very high beam energies using e.g. plasma acceleration,
to very high luminosities through the use of energy-recovery linac technology,
or a photon-photon collider option).
These discussions are now underway.
The choice of L* is in principle more flexible, and may evolve with facility upgrades.
We may assume that it will be similar to that used in the ILC (4.1~m) or CLIC (6~m) designs.
The MDI will be influenced by the main linac technology, since the spacing between bunches is typically much shorter in
normal-conducting than super-conducting accelerators. As an example, the shorter inter-bunch distance at CLIC compared to ILC requires
a larger crossing angle to minimise interactions between outgoing bunches with the incoming bunches of the other beam.

\section{Collision rate} \label{sec3}

A circular collider, such the FCCee, will provide the detectors with a quasi-continuous stream of bunch collisions. The FCCee
is designed to operate at constant synchrotron power loss at different CM energies, altering the average spacing between
bunches at different energy stages, from $\sim$30~ns at 91~GeV to $\sim 5~\mu$s at 365~GeV. 
On the other hand, collisions at linear colliders are arranged in ``trains'', separated by a relatively long
quiet time with no collisions (e.g. at ILC, 1312 bunch collisions each separated by 554~ns occur within around 1~ms,
followed by a quiet period of 199~ms until the arrival of the next train.

The ILC detector designs have made use of this bunch train operational mode to save power and cooling requirements
by by reading out data and turning off parts of the front-end electronics during the quiet periods.
This enables the power, data flow and cooling infrastructure within the detector to be dimensioned to cope with
the requirements averaged over both active and quite periods.
Such service requirements are therefore much lower than those in continuous operation of FCC-ee.

\section{Magnetic fields}

The strength of the experiments' solenoid field affects several aspects of the detector's performance:
\begin{itemize}
\item
  resolution of the magnetic spectrometer system (tracker + field) used to measure the transverse momentum of charged particles;
\item
  the performance of particle flow. The field encourages separation of charged and neutral particles at the calorimeter,
  reducing the potential for ``confusion'' between charged and neutral energy deposits; and
\item
  the constraining force it exerts on the vast numbers of low ${\textrm p_T}$ electrons and positrons produced in
  beamstrahlung,
preventing almost all of them from entering the beampipe and tracking system (e.g. Vertex Detector), where they would
otherwise swamp the detector with background.
\end{itemize}
The field also constrains the motion of ionisation electrons produced in
the TPC tracker envisaged for ILD, helping attain the required spatial resolution.

At linear colliders, at which beam bunches are discarded after use, there is freedom to choose 
essentially any detector solenoid strength. The field strengths chosen for the ILC detectors are 3.5~T (ILD) and 5~T (SiD).
At circular colliders individual bunches are re-circulated many times, so care must be taken
not to unduly disturb the beams and increase the emittance.
The combination of the beams' crossing angle and the detector's solenoid field results in a net
horizontal field component seen by the beams, inducing a vertical kick to the beam.
This kick must be compensated, and FCCee optics include ``compensating solenoids'' for this reason.
In the standard ``local compensation'' scheme, a 5~T magnet is placed just in front of the
last quadrupoles, and the detector field is required to be no stronger than 2~T to maintain the beam emittance.
If an alternative ``non-local compensation'' scheme is used, in which the compensating solenoids are placed outside the
detector volume, this solenoid field limit may be slightly relaxed.

\section{Machine backgrounds}

The different beam parameters and MDI designs at different colliders can have an important 
influence on the detector backgrounds. Different bunch dimensions and populations give different
amounts of beamstrahlung electrons and positrons. These are constrained to a region around the
beampipe by the magnetic fields. A very small fraction of them may directly hit the beampipe and vertex detector,
while the remainder interact with material in the forward region, typically associated with the
MDI. Photons produced in these interactions (including a major contribution of 511~keV photons from positron annihilation)
can easily traverse the thin inner regions of the detector, inducing hits in sensitive detectors.


The interaction of the beamstrahlung particles with detector material also has a strong influence
on the level of visible backgrounds. At ILC, with its long L*, the first material which most
beamstrahlung particles come across is the BeamCal, at z=4~m, deep inside the forward calorimeter system.
This region is quite well shielded from the central tracking region by the surrounding calorimeters.
The situation at FCCee is different, with the beampipe crotch at z=1.3~m being the primary scattering source.
This crotch is less well shielded from the central detector region.

Studies have shown that the amount of beamstrahlung background seen by e.g. ILD's TPC per bunch crossing is similar 
at ILC and FCC-ee, with the competing effects of the greater bunch focus at ILC and the more intrusive MDI at FCC-ee
more or less compensating each other \cite{Jeans:2024nba}. For subdetectors which have a long integration time,
in particular the TPC with its maximum ion drift time of around half a second, the much higher collision
rate at FCC-ee becomes a critical factor.

Synchrotron radiation from the upstream beamline and final focus system is also expected to be a major source of background;
these sources are now being studied.

\section{Subdetector technologies and integration}

The bunch train operation typical at linear colliders permits the use of ``power pulsing'', in which
power-hungry parts of the front-end electronics are powered up only during collisions, and
powered down in the relatively long periods between bunch-train arrivals.
Data is stored locally during collisions, and read out at leisure during the following quiet period.
In the case of ILC, collisions occur during a period of around 1~ms, at a frequency of 5~Hz:
a nominal duty factor of 0.5\%. The use of power pulsing drastically reduces subdetectors' need for
power, cooling, and high-speed data transfer. This results in a low mass system to
supply these services to sensor elements, with the consequent advantages of a lightweight
and transparent tracking detector.

Moving to a circular collider with essentially continuous collisions precludes the use of
power pulsing. A reasonable design must balance the desired detector granularity, the
power budget for front-end electronics, and the effects of the material budget due to services
on measurement resolution through multiple scattering and material interactions.
These considerations may well result in e.g. calorimeters with less sampling layers and
larger readout granularity, with more massive active cooling circuits.

\section{Current status in ILD}

ILD is currently adapting its concept for application to the FCC-ee.
The key unavoidable difference with respect to ILC is the design of the MDI system
and the limited strength of the experiment's magnetic field.
To make first investigations of the effect of these changes, new simulation models of ILD have been
developed which apply these changes, while keeping as much as possible of the
detector unchanged.

\begin{figure}
  \center
  \includegraphics[width=0.49\textwidth]{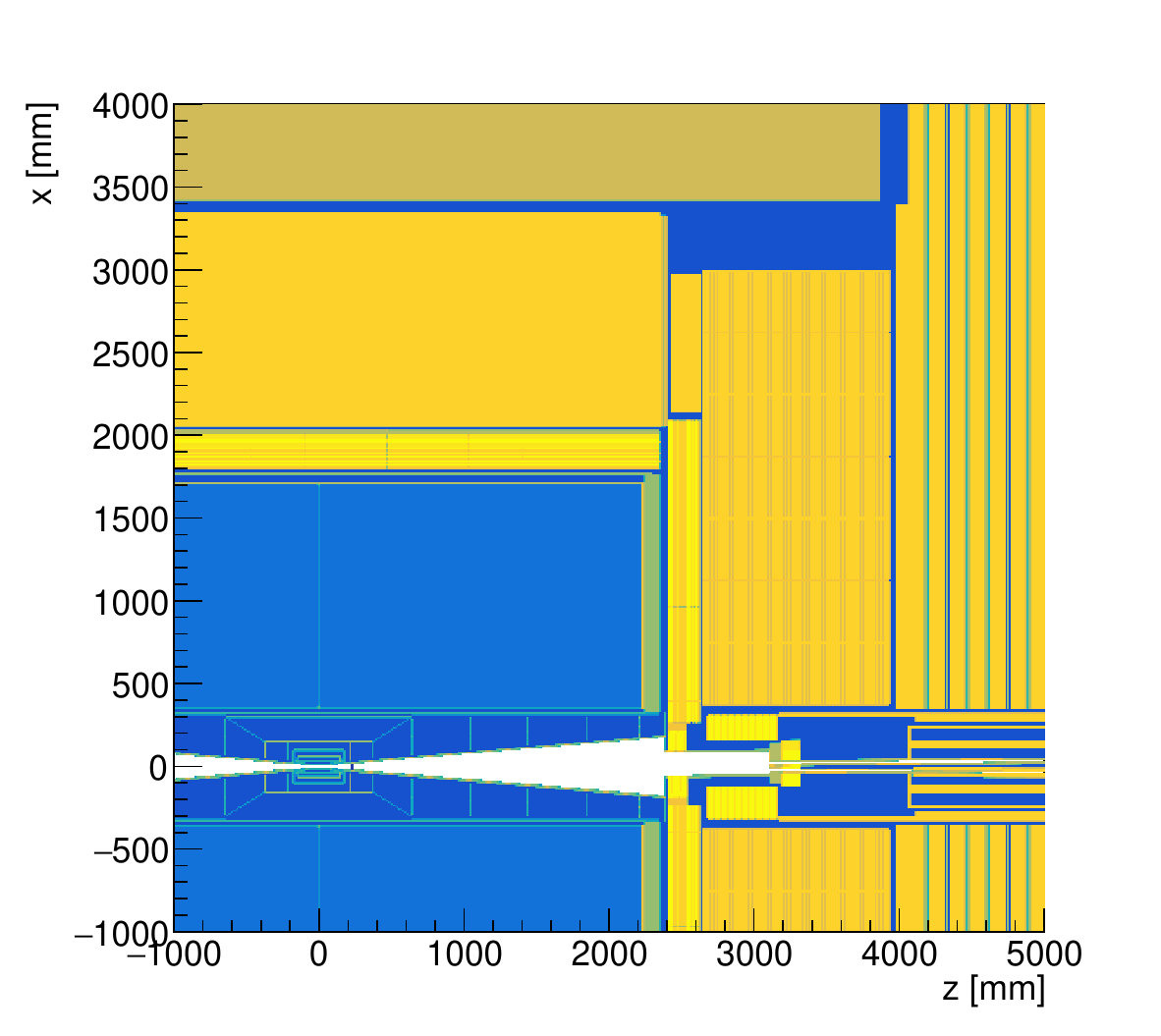}
  \includegraphics[width=0.49\textwidth]{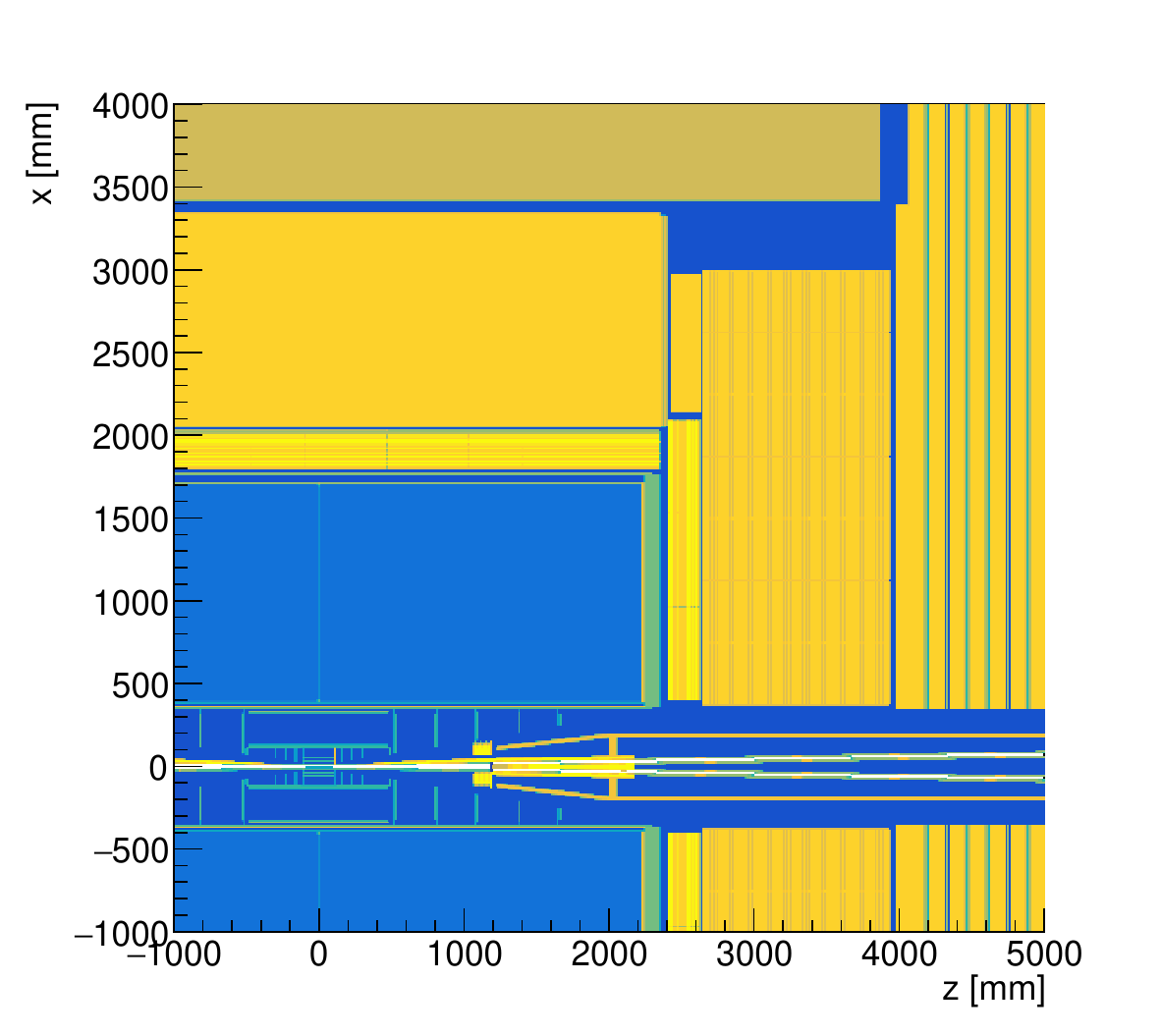}
  \caption{Comparison of the ILD simulation models for ILC (ILD\_l5\_v02) and FCC-ee (ILD\_FCCee\_v01), showing the change in the
    inner detectors and MDI system. The TPC and main calorimeters are left unchanged, while the nominal magnetic field strength has
    been reduced from 3.5 to 2~T.}
  \label{fig:detPics}
\end{figure}

Defining these models has profited from the plug-and-play nature of the
detector descriptions in the k4geo package, which makes it easy to incorporate
a common description of the MDI in ILD, and to adapt the design of the inner silicon tracking system
developed by the CLD group.
Cross-sections of the ILD simulation models for ILC and FCC-ee are shown in Fig.~\ref{fig:detPics}.

Detector subsystems are also affected by the different FCCee environment. We give a few thoughts on two particular
systems, the TPC and high granularity calorimeters which are envisioned for use in ILD.

\subsection{Time Projection Chamber}

Perhaps the most critical and unique question for ILD to answer is whether the Time Projection Chamber (TPC) can function effectively at FCC-ee,
particularly at the high luminosity Z-pole run.
Of all subdetector technologies currently proposed for FCC-ee detectors, the TPC is in some respects the slowest.
Ions drift slowly within its gas volume (expected speed around 5~m/s), over a maximum drift length of 2.2~m,
giving a maximum drift time of around half a second.
The TPC volume becomes filled with a sea of ions, contributed to by collisions which occurred over the previous
half-second.
The major concerns are the arrival rate of background-induced ionisation electrons at the readout plane and the electric field
distortions induced by the ion cloud. At ILC and LCF, with the modest bunch crossing rate
(integrating a few thousand bunch crossings over the half-second maximum ion drift time), and the possibility
to block secondary ions using a gating device during the inter-train period, the distortions have been shown to
be $\mathcal{O}(10 \mu m)$, smaller than the single hit position resolution.
At a circular collider Z-pole run, half a second corresponds to $\mathcal{O}(10^7)$
bunch crossings and no active gating is possible, so distortions are expected to be larger by orders of magnitude.

We are currently estimating the background levels due to various machine-related processes within the TPC,
using simulated data samples provided by the FCCee MDI group. Preliminary results presented at this
workshop~\cite{Watanabe} show that beamstrahlung at the FCCee-91 can distort reconstructed
particle trajectories by up to a few cm at the inner TPC radius, with typical distortions of a few mm
in the main body of the TPC. These distortions seem rather stable with time, suggesting that they
can be effectively corrected.

Predictions of background levels due to synchrotron radiation seem to depend strongly on the MDI design,
particularly collimators, masks, and shielding, as well as the precision with which accelerator elements are aligned.
Efforts to understand, reduce, and mitigate background levels are in progress.

The high expected hit rates suggest that a pad-based readout technologies will suffer from extremely high occupancies,
which argues for the use of a TPC with pixel-based readout, as being developed within the LC-TPC collaboration~\cite{vanBeuzekom:2025ias}.

\subsection{Calorimeters}

The high granularity calorimeters being considered for detectors at ILC and their descendants at FCCee
have a highly integrated design, with front end electronics placed within the calorimeter volume.
The large channel count results in non-negligible dissipation of power within the calorimeters
during operation. At ILC, the use of power pulsing significantly reduces the amount of heat produced,
and allows it to to passively guided to the edge of detector modules, where it is absorbed by an actively cooled heat sink.
This limits temperature differences within the calorimeter modules to around 10~K.

With the change to continuous running, such an approach will no longer work.
Solutions in which cooling pipes are integrated into the absorber structure are now being studied,
inspired by the similar approach taken in the CMS HGCAL \cite{siwecalCool, ahcalCool},
at the cost of less longitudinal granularity due to thicker layers. A reduction in readout granularity
within layers can also help reduce the cooling needs. The optimal balance between longitudinal
and transverse granularity should be identified.

\section{Outlook}

Detectors at the ILC and circular Higgs Factory colliders such as the FCCee share many physics goals, and therefore many
performance requirements. However, there are several important differences regarding the
environment in which the detectors must operate.

We have a good general idea of the aspects of ILC detectors which should be adjusted to operate at
circular electron-positron Higgs factories: technologies able to cope with high rates,
increased power and cooling needs, and limitations on the detector field.
Some of these aspects still need to be understood more precisely.
In particular the detector backgrounds due to beam-related processes such as beamstrahlung and
synchrotron radiation, which depend strongly on the details of the
still developing Machine-Detector Interface design, still need robust estimates
in order to make some of the important design choices to define the detectors for FCCee.


\section*{Acknowledgements}

I thank my colleagues in SiD and ILD for many useful discussions,
and the organisers of the LCWS2025 conference for their invitation to contribute to the proceedings.

\end{document}